\documentclass[12pt,titlepage]{article}
\usepackage{graphicx}
\begin{document}

\title{Spin-isospin model of NaV$_2$O$_5$}

\author{M. V. Mostovoy, J. Knoester, and D. I. Khomskii\\ \\
{\em Theoretical Physics Institute and Materials Science
Center,}\\
{\em University of Groningen, Nijenborgh 4,}\\
{\em 9747 AG Groningen, The Netherlands}}

\begin{abstract}

We argue that in the quarter-filled ladder compound
NaV$_2$O$_5$ the quasi-one-dimensional spin system is
strongly coupled to a low-energy antiferroelectric mode
of the excitonic type. This mode results from the
interplay between the electron hopping along the rungs
of the vanadium two-leg ladders and the Coulomb
repulsion between electrons. The charge ordering
observed in sodium vanadate at $T_c = 34$K corresponds
to the softening of the antiferroelectric mode. We
consider the spin-isospin model, which describes the
spin and low-energy charge degrees of freedom in sodium
vanadate. Within this model we explain the observed
anomalous temperature-dependence of the dielectric
susceptibility at $T_c$ and the midinfrared absorption
continuum. We identify the broad structure in the
low-energy optical absorption spectrum of NaV$_2$O$_5$
with the continuum formed by two spinons and one
low-energy charge excitation.

\end{abstract}

\maketitle

\section{Introduction}
\label{introduction}

Sodium vanadate (NaV$_2$O$_5$), first studied in the
seventies \cite{Carpy}, has become an object of
intensive experimental and theoretical investigations
after the discovery of the phase transition at $T_c
=34$K \cite{Isobe}. Several physical concepts, such as
bipolarons \cite{Chakra}, spin-Peierls transition
\cite{Isobe,Fujii}, and charge ordering
\cite{Ohama,Fukuseo,KM}, have been proposed to explain
the properties of this material. Yet, despite all the
efforts, the low-temperature crystal structure of
sodium vanadate remains controversial and the nature of
the low-energy excitations in this material is still
poorly understood.

\begin{figure}
\centering
\includegraphics[width=7cm]{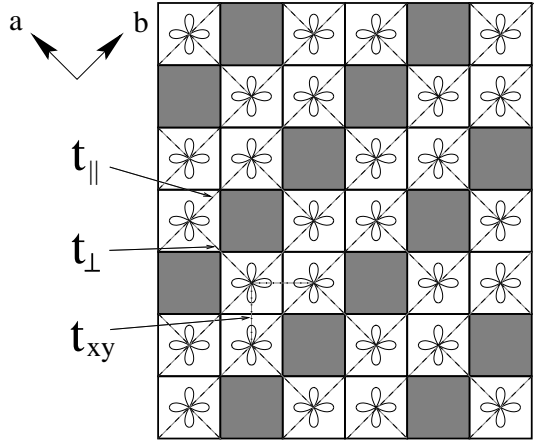}
\caption{
\footnotesize
The crystal structure of the V-O plane in NaV$_2$O$_5$: oxygens
are located at the corners of the plaquettes, while V ions are
located at their centers; the shaded plaquettes are vacant.  Also
shown are the vanadium ladders (dashed lines) and the relevant
$d_{xy}$-orbitals of the V ions.}
\label{VOlayer}
\end{figure}

In $\alpha^{\prime}$-NaV$_2$O$_5$ the vanadium ions
form two-leg ladders organized in layers (see
Fig.~\ref{VOlayer}).  Until recently it was assumed
that the vanadium ladders with one electron per rung
are equivalent to spin-$\frac{1}{2}$ chains, so that
sodium vanadate is a quasi-one-dimensional spin system.
Indeed, this material is a good insulator and its
magnetic susceptibility has temperature behavior
similar to that of the Heisenberg spin-$\frac{1}{2}$
chain with the exchange constant $J \sim 560$K
\cite{Hemberger}. At $T_c = 34$K sodium vanadate
undergoes a phase transition into a state with a spin
gap.  Below $T_c$ the lattice period along the ladders
($b$-direction) and perpendicular to ladders within the
layers ($a$-direction) doubles, while the period in the
direction perpendicular to the layers ($c$-direction)
increases by a factor of four \cite{Fujii}. The
transition was initially interpreted as a spin-Peierls
transition, driven by the instability of
spin-$\frac{1}{2}$ chains against the formation of
local dimers.

However, more recent experimental studies of NaV$_2$O$_5$ revealed
some inconsistencies in that interpretation.  First, sodium
vanadate does not show the strong suppression of $T_c$ by magnetic
field, characteristic for spin-Peierls systems
\cite{Kremer,Buechner}.  Second, the entropy released at the
transition is considerably higher than the expected release of the
spin entropy \cite{Hemberger,Kremer,Buechner}.  Third, the
dielectric susceptibility, measured by microwave absorption, has
an anomaly at the transition temperature, which was not observed
in the inorganic spin-Peierls material CuGeO$_3$ \cite{Smirnov}.

Recent X-ray \cite{Meetsma,Schnering,Smolinski}, as
well as NMR measurements \cite{Ohama}, show that above
$T_c$ all vanadium ions are equivalent (V$^{4.5+}$),
whereas below $T_c$, a charge disproportionation
occurs. There is, however, still a controversy about
the number of distinct V sites and their arrangement in
the low-temperature phase
\cite{Ohama,Ludecke,deBoer,Palstra}.

It was then suggested that the driving force of this
transition is the charge ordering of electrons in the
quarter-filled vanadium ladders and that the spin-gap
opens due to alternation of the spin-exchange
constants, resulting from the charge ordering
\cite{Fukuseo,KM}. Since sodium vanadate is an
insulator both above and below $T_c$, the charge
ordering in this material does not originate from the
Fermi surface instability, as in charge-density-wave
materials. Therefore in Ref. \cite{KM} we proposed a
spin-isospin model, in which the rungs of the vanadium
ladders are assumed to be predominantly occupied by one
electron. The isospins describe the low-energy charge
degrees of freedom, which in sodium vanadate are
electric dipoles located on the rungs of the vanadium
ladders. The isospin Hamiltonian (Ising model in a
transverse field) is identical to the Hamiltonian used
to describe (anti)ferroelectric transitions
\cite{Blinc}. Below $T_c$ the dipoles become ordered.
Their directions alternate along the ladders
\cite{Fukuseo,KM}, which corresponding to a zigzag
ordering of charges. The antiferroelectric nature of
charge displacements in the vanadium-oxygen layers was
confirmed by the observation of anomalies in the
dielectric susceptibility close to $T_c$ \cite{Smirnov,
Sekine, Poirier}. In Ref. \cite{KM} we showed that the
antiferroelectric ordering in the system of vanadium
ladders opens a spin gap.

We recall that, depending on whether the variable
associated with the dipole moment of a unit cell is
quantum or classical, there are two limiting types of
(anti)ferroelectric transitions: a displacive
transition and an order-disorder transition
\cite{Blinc}. If the tunneling between the two states
with different orientations of the dipole moment can be
neglected, the transition is of the order-disorder
type, described by a classical Ising model. If, on the
other hand, the hopping between the two states is
non-negligible, the transition is described by the
quantum Ising model in a transverse field. The quantum
dynamics of the electric dipoles gives rise to a branch
of excitations that correspond to dipole flips
propagating from one unit cell to another. At the
transition temperature these excitations become
gapless. The softening of the isospin excitation occurs
at the wave vector of the superlattice structure
appearing in the ordered phase.

In sodium vanadate the amplitude of the electron
hopping along the rungs of the vanadium ladders, which
plays the role of the tunneling amplitude between two
states of the electric dipole, is of the same order as
the Coulomb interaction between electrons on
neighboring rungs (1eV) that is responsible for the
charge ordering. Thus the dynamics of the
antiferroelectric mode in sodium vanadate is
essentially quantum and, at least close to the
transition temperature, there should be low-energy
charge excitations in this material.

For the antiferroelectric ordering (see Sec.~\ref{CO})
the soft mode has a nonzero wave vector and, therefore,
cannot be directly excited in optical absorption and
Raman scattering above the transition temperature.
Nevertheless, we show in this paper that the optical
data do provide an evidence for the presence of the
low-energy charge excitations in sodium vanadate.

The optical spectrum of sodium vanadate contains the
broad absorption band covering almost the whole
midinfrared region of frequencies
\cite{Damascelli,Popova99}. A broad peak was observed
also in Raman experiments
\cite{Popova99,Golubchik,Fisher}. It was suggested in
Ref. \cite{Damascelli}, that the low-energy absorption
is due to the photoexcitation of the two-spinon
continuum. However, to convert a photon into a pair of
spinons, the presence of a permanent electric dipole
moment is required (both above and below $T_c$, since
the continuum is observed at all temperatures). This
would be possible if sodium vanadate would have the
ferroelectric chain-like structure that was originally
proposed for this material \cite{Carpy}, but which
later was found to be inconsistent with X-ray data
\cite{Meetsma,Schnering,Smolinski}.

In this paper we interpret the broad peak in the
low-frequency optical absorption spectrum as a
three-particle continuum: two spinons plus the
low-energy antiferroelectric excitation. Our mechanism
crucially depends both on the presence of the soft
mode, associated with the charge ordering, and on the
coupling of this mode to the spin excitations.

This coupling naturally arises in the spin-isospin
model discussed in Sec.~\ref{spin-isospin}. In
Sec.~\ref{CO} we consider the charge ordering in this
model, which, as was already mentioned, is related to
the softening of the isospin excitations discussed in
Sec.~\ref{excitons}. In Sec.~\ref{anomaly} we apply our
model to explain the anomaly of the dielectric
susceptibility at $T_c$, while in
Sec.~\ref{oabsorption} we show how the coupling between
spin and isospin excitations gives rise to the
low-energy optical absorption. We discuss our results
in the concluding Sec.~\ref{disclusions}.

\section{Spin-isospin model}
\label{spin-isospin}

The experimental observations mentioned in
Sec.~\ref{introduction} indicate that the
low-temperature properties of sodium vanadate cannot be
described by only considering the spin and lattice
dynamics, but that it is necessary to include also the
charge degrees of freedom. On the other hand, sodium
vanadate is known to be a good insulator, both above
and below $T_c$. We, therefore, argued in \cite{KM}
that the relevant electronic excitations in this
material are excitons. In this section we briefly
discuss the ``spin-isospin'' model, introduced in
Ref.\cite{KM}, which describes the low-energy
excitations in sodium vanadate.

In that model we restrict ourselves to the states with
only one electron per rung of the vanadium ladder. This
was first suggested by Smolinski {\em et al.} on the
basis of LDA calculations \cite{Smolinski}, which show
that the electron hopping amplitude along rungs,
$t_{\perp}$, is significantly higher than the
amplitudes of hopping between the sites of different
rungs (of which the largest is the hopping amplitude
along ladders, $t_{\parallel}$). If one first neglects
the interrung hopping and the interrung
electron-electron interactions, then on each rung of
the quarter-filled vanadium ladder there is one
electron occupying the symmetric bonding state with
energy $-t_{\perp}$.  Due to the large value of the
on-site Coulomb repulsion (which we assume to be
infinite), an electron can hop only to an empty site of
a neighboring rung. Then the hopping to a neighboring
rung results in an energy increase $2 t_{\perp}$. Thus,
the quarter-filled two-leg ladder becomes equivalent to
the half-filled Hubbard chain with an effective
``on-rung'' repulsion $U_r = 2 t_{\perp}$.

This explains, in principle, why vanadium ladders are
insulating. However, for the values of the hopping
amplitudes, obtained in \cite{Smolinski}, $t_{\perp} =
0.38$eV and $t_{\parallel}=0.17$eV, the width of the
one-dimensional band, $4 t_{\parallel}$, is comparable
with $U_r$. Thus the charge fluctuations on the ladder
rungs are not small and the treatment of the vanadium
ladder as a spin chain, as it was done in
\cite{Smolinski}, is not well-justified.

The situation improves if we take into account the
Coulomb interactions between electrons on different
rungs. The latter lead to an increase of the value of
the ``on-rung'' repulsion. For instance, if we include
the repulsion between the nearest-neighbor sites in
ladders, $V$, then $U_r \approx 2t_{\perp} +
\frac{V}{2}$, for small $V$. In general, the interrung
Coulomb interactions make $U_r$ not a well-defined
quantity, as it becomes dependent on the positions of
many electrons.

Thus, on the one hand, the interrung Coulomb interactions help
to justify the assumption of one electron per rung. On the other
hand, they strongly mix the symmetric and antisymmetric states
on each rung, as the typical value of such interactions is of
the same order as the energy separation between these two
states, $2 t_{\perp} \approx 0.75$eV. Thus, the electron
position on a rung becomes an important additional degree of
freedom, which makes the vanadium ladder different from a spin
chain.

\begin{figure}
\centering
\includegraphics[width=6cm]{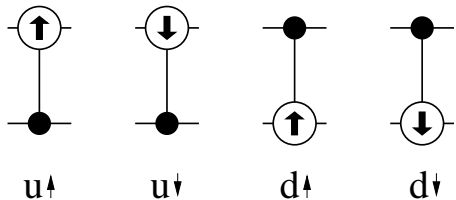}
\caption{ \footnotesize The four different states of a
single electron on a rung represented as the
eigenstates of the spin $S = \frac12$ and the isospin
$T = \frac12$ operators.} \label{fourstates}
\end{figure}

\begin{figure}
\centering
\includegraphics[width=12cm]{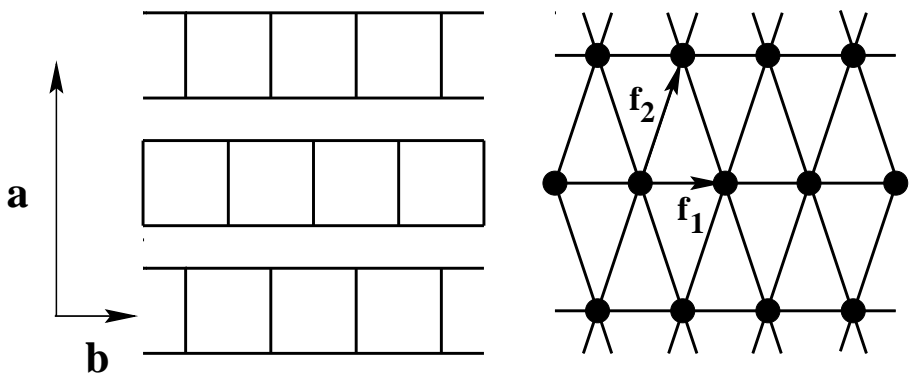}
\caption{ \footnotesize From vanadium ladders
(left-hand side of the picture) to the effective
lattice (right-hand side of the picture). The sites of
the latter should be identified with the centers of the
rungs of the V ladders.} \label{efflat}
\end{figure}

The two states, $u_{\bf n}$ and $d_{\bf n}$,
corresponding to the two possible positions of a single
electron on the rung ${\bf n}$, can be described as the
up and down eigenstates of an isospin-$\frac{1}{2}$
operator, $T_{\bf n}^z$. With two spin projections
there are in total four different states of an electron
on a rung (see Fig.~\ref{fourstates}).  The spin ${\bf
S}_{\bf n}$ and the isospin ${\bf T}_{\bf n}$ are
defined on the lattice, the sites of which correspond
to the rungs of vanadium ladders (see
Fig.~\ref{efflat}).  In this paper we ignore the
three-dimensional structure of sodium vanadate and only
consider one oxygen-vanadium layer.  Then the lattice
of our model is triangular.  Furthermore, for our
considerations it will not be important that in each
layer the orientations of pyramids in the
nearest-neighbor ladders are opposite to each other
(which is why the unit cell above $T_c$ contains two
ladders). Therefore, we chose ${\bf f}_1 = {\bf b}$ and
${\bf f}_2 = \frac{1}{2}({\bf a} + {\bf b})$ as the
basis of the unit cell in the high-temperature phase.

The charge part of the Hamiltonian of the model, which
only includes the isospin operators, has the form:
\begin{equation}
H_T  =  -  2 t_{\perp} \sum_{{\bf n}} T^x_{{\bf n}} +
\frac{1}{2} \sum_{\bf nm}  V_{\bf nm}
 \, T^z_{n} \, T^z_{m},
\label{HT}
\end{equation}
where $t_{\perp}$ is the amplitude of electron hopping along the
ladder rungs and the second term describes the Coulomb repulsion
between electrons on different rungs.  The amplitudes $V_{\bf
nm}$ are defined by
\begin{equation}\label{Isingint}
V_{\bf nm} = V(u_{\bf n},u_{\bf m}) + V(d_{\bf n},d_{\bf m}) -
V(u_{\bf n},d_{\bf m}) - V(d_{\bf n},u_{\bf m}), \label{Vnm}
\end{equation}
where, {\em e.g.}, $V(u_{\bf n},d_{\bf m})$ is the
energy of the Coulomb interaction between an electron
on rung ${\bf n}$ in the up state and an electron on
rung ${\bf m}$ in the down state. The Coulomb
interactions favor a certain charge ordering, which in
our model corresponds to an ordering of the
$z$-projections of the isospins, while the electron
hopping counteracts the ordering. Since the hopping
term does not commute with the Coulomb term, the
isospins are quantum (rather than classical) degrees of
freedom, like the orbital degrees of freedom in Mott
insulators with orbital degeneracy \cite{KK}.

We now turn to the spin contributions to the
Hamiltonian. The spin exchange results from virtual
hopping of electrons between neighboring rungs, and the
exchange interaction is the strongest in the ladder
direction. For the half-filled Hubbard chain the
exchange interactions between spins on two neighboring
sites has the form $\frac{2t^2}{U} S_{12}$, where
$S_{12} = 2 \left(\bf{S}_1 \cdot \bf{S}_2 \right) +
\frac{1}{2}$ is the operator that exchanges $\bf{S}_1$
with $\bf{S}_2$. In the two-leg ladder it is possible
to have in the intermediate state a doubly occupied
rung without doubly occupied sites. Thus even for
infinite $U$ the exchange coupling is nonzero and $U$
has to be replaced by the ``on-rung'' repulsion $U_r$.
In the infinite $U$ case the exchange of spins
necessarily involves the exchange of the
$z$-projections of the isospins on two neighboring
rungs ($u_1d_2 \leftrightarrow d_1u_2$), so the
spin-exchange interaction has the form:
$\frac{2t^2}{U_r} S_{12} \left(T_1^{+} T_2^{-} +
T_1^{-} T_2^{+}\right)$. Thus, instead of a pure
spin-exchange Hamiltonian we obtain a spin-isospin
interaction:
\begin{equation}
H_{ST}^{(1)} = A\!  \sum_{\bf n}\!  \left({\bf S}_{\bf n}
\!\cdot\!  {\bf S}_{{\bf n} + {\bf f}_1}\right) \left(T^x_{\bf n}
T^x_{{\bf n}+{\bf f}_1} \!+\!  T^y_{\bf n} T^y_{{\bf n} + {\bf
f}_1} \right).
\label{HST1}
\end{equation}
Here, the rungs ${\bf n}$ and ${\bf n} +{\bf f}_1$ are the two
nearest-neighbor rungs of the ladder (see Fig.~\ref{efflat}) and
the interaction amplitude $A = \frac{8 t_{\parallel}^2}{U_r}$,
$U_r$ being the effective ``on-rung'' Coulomb repulsion discussed
above.

We note that Eq.(\ref{HST1}) gives only an approximate
description of the spin-isospin interaction in sodium
vanadate, because in the presence of interrung Coulomb
interactions, both initial and intermediate states in
the spin-exchange process are complicated many-electron
states, so that the increase of the energy in the
virtual state cannot be described by a single constant
$U_r$. However Eq.(\ref{HST1}) contains many basic
ingredients, necessary for a qualitative description of
the properties of sodium vanadate.

In particular, treating the spin-isospin interaction in the
mean-field approximation we obtain the effective spin-exchange
coupling constant:
\begin{equation}
J  = A \langle T^x_{\bf n} T^x_{{\bf n}+{\bf f}_1} + T^y_{\bf n}
T^y_{{\bf n}+{\bf f}_1} \rangle,
\end{equation}
where the brackets denote the thermal and quantum
average of the isospin operators.  The exchange
coupling is, therefore, temperature-dependent, which
may account for the deviation of the magnetic
susceptibility of sodium vanadate from the
Bonner-Fisher curve \cite{Buechner, Johnston}.  Using
the Hamiltonian Eq.(\ref{HT}), one can show that, above
$T_c$, the exchange coupling increases as the
temperature decreases, which is in agreement with the
conclusion of Ref.~\cite{Johnston}, where such an
increase was introduced phenomenologically in order to
reconcile the magnetic susceptibility data for
NaV$_2$O$_5$ with the temperature dependence of the
magnetic susceptibility of the Heisenberg
spin-$\frac{1}{2}$ chain.

However, there is one very important effect of the
electronic charge distribution on the spin exchange
that is not described by Eq.(\ref{HST1}). Namely, the
Hamiltonian Eq.(\ref{HST1}) does not explain the
spin-gap opening due to the zigzag charge ordering. In
order to find the term in the spin-isospin interaction
responsible for the spin-gap opening, we note that the
exchange between spins on rungs ${\bf n}$ and ${\bf
n}+{\bf f}_1$ depends on the position of electrons in
the nearest rungs of the two neighboring ladders, ${\bf
n}+{\bf f}_2$ and ${\bf n}+{\bf f}_1-{\bf f}_2$ (see
Fig.~(\ref{efflat})). Roughly speaking, electrons
affect the hopping amplitudes between the rungs of
nearest-neighbor ladders, which in turn, affect the
spin-exchange. The corresponding Hamiltonian describing
such an influence has the form:
\begin{equation}
H_{ST}^{(2)}  = B \sum_{\bf n}\! \left(T^z_{{\bf n}+{\bf f}_2} -
T^z_{{\bf n}+{\bf f}_1-{\bf f}_2}\right) \left({\bf S}_{\bf n}
\!\cdot\! {\bf S}_{{\bf n}+{\bf f}_1}\right). \label{HST2}
\end{equation}
As was argued in Ref.~\cite{KM}, this interaction is
responsible for the opening of the spin gap in the
charge ordered phase. Indeed, for the zigzag structure
with a doubling of the period in the $a$ and $b$
directions (see Fig.~{\ref{costates}a), the exchange
alternates along ladders, which opens a spin gap.

To end this section, we note that the virtual hopping
of electrons on neighboring rungs, apart from affecting
the spin-exchange, also results in corrections to the
pure isospin Hamiltonian Eq.(\ref{HT}), which we
neglect here, as they are relatively small.
Furthermore, for large, but finite on-site Hubbard $U$,
one obtains also the spin-isospin interaction term of
the form:
\begin{equation}
H_{ST}^{(3)} = A^{\prime}\!  \sum_{\bf n}\!
 \left({\bf S}_{\bf n} \!\cdot\!  {\bf S}_{{\bf n}
 + {\bf f}_1}\right) T^z_{\bf n} T^z_{{\bf n}+{\bf f}_1},
\label{HST3}
\end{equation}
where $A^{\prime} \propto t_{\parallel}^2 / U$.

\section{Charge ordering}
\label{CO}

In this section we ignore spins and discuss the phase
diagram for the Hamiltonian (\ref{HT}), which describes
the charge dynamics. This is justified as, according to
our assumption, the driving force of the transition in
sodium vanadate is the electronic charge ordering.

Since the Hamiltonian (\ref{HT}) is real, the average
value of $T_{\bf n}^y$ is zero at all temperatures
(unless the time-reversal symmetry is spontaneously
broken in the ordered state).  On the other hand,
$M_{\bf n}^x = \langle T_{\bf n}^x \rangle \neq 0$ at
all temperatures, due to the presence of the
``external'' transverse field $2 t_{\perp}$.  The
ordered state of the isospin system is characterized by
a nonzero value of $M^z_{\bf n} = \langle T_{\bf n}^z
\rangle$.  The ordering of isospins in the
$z$-direction corresponds to the modulation of the
electronic density on the rungs of the vanadium
ladders.  The type of charge ordering depends on the
details of the Coulomb interactions $V_{\bf nm}$.

\begin{figure}
\centering
\includegraphics[width=9cm]{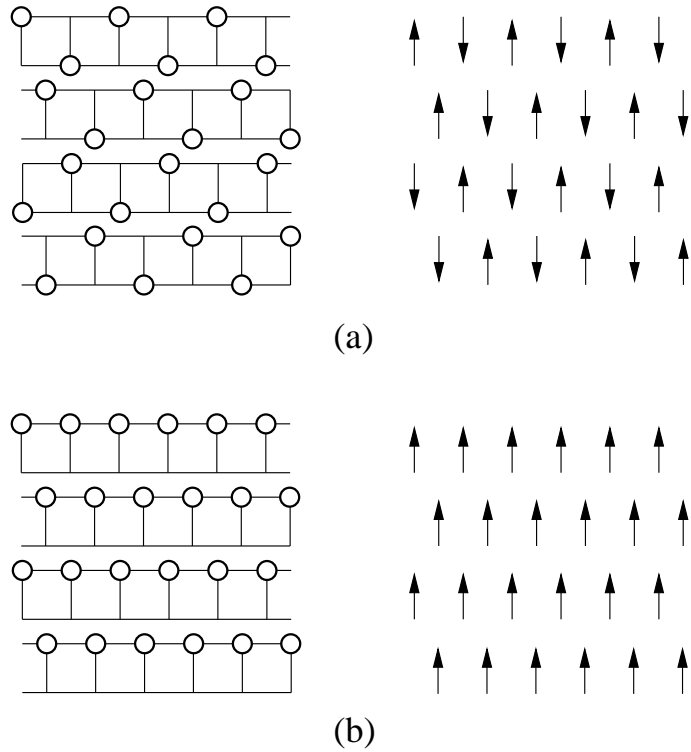}
\caption{ \footnotesize Two charge ordered states:
zigzag (or antiferroelectric) ordering with the
doubling of the unit cell in the directions $a$ and $b$
(a) and the chain-like (ferroelectric) structure (b).
The left part of each picture shows the distribution of
the electronic density, while the right part shows the
corresponding ordering of isospins.} \label{costates}
\end{figure}

If we use as estimates the unscreened Coulomb
interactions between electrons, then, according to
Eq.(\ref{Isingint}), the Ising interaction between the
isospins on the nearest-neighbor rungs in the ladder,
$J_A = \frac{1}{4} V_{{\bf n},{\bf n} + {\bf f}_1} =
0.62 $eV. This ``antiferromagnetic'' interaction favors
a zigzag structure, like {\em e.g.} the one shown in
Fig.~\ref{costates}a. On the other hand, the
interaction between the isospins on the nearest rungs
of two neighboring ladders, $J_F = \frac{1}{4} V_{{\bf
n},{\bf n} + {\bf f}_2} = 0.54$eV, is ``ferromagnetic''
(for simplicity, we neglect the shifts of vanadium ions
out of the basal plane). The ``ferromagnetic''
interaction favors the structure shown in
Fig.~\ref{costates}b, in which electrons occupy with
higher probability one chain of each ladder. In
general, the charge ordering corresponds to the
ordering of the isospins in the $z$-direction:
\begin{equation}\label{odpar}
\langle T^z_{\bf n} \rangle = e^{i {\bf Q} \cdot {\bf x}_{\bf n}}
M^z,
\end{equation}
where ${\bf x}_{\bf n}$ is the position of rung ${\bf
n}$, and ${\bf Q} = (\frac{\pi}{f_1},\frac{\pi}{f_2})$
for the zigzag structure, while for the chain structure
${\bf Q} = 0$.

If one would take into account only the Ising
interactions between the neighboring rungs, then for
$J_A > |J_F|$ the energy of the zigzag structure is
lower energy than that of the chain structure. However,
since the difference between $J_A$ and $|J_F|$ is
relatively small, the effects of screening of the
Coulomb interactions and the long-range interactions
between the isospins are important for the
determination of the nature of the charge-ordered
state.

Though the precise calculation of the Ising
interactions between the iso\-spins is very difficult,
one can discriminate between different charge-ordered
structures on the basis of the available experimental
data. In particular, the chain-like ordering (see
Fig.~\ref{costates}b) does not result in the increase
of the lattice periodicity in the $a$ and $b$
directions observed in NaV$_2$O$_5$ below $T_c$.
Furthermore, the chain structure corresponds to a
ferroelectric state. As we shall show below, the
optical response of such a system would be quite
different from that of sodium vanadate. On the other
hand, the zigzag charge ordering results in a doubling
of the lattice period in the $b$-direction. Moreover,
if the zigzags in the next-nearest-neighbor ladders
have opposite phases (as in Fig.~\ref{costates}a), then
the lattice period in the $a$-direction also doubles.

The generic phase diagram of the Ising model in a
transverse field is shown in Fig.~\ref{diagram}.  For
large values of $t_{\perp}$ the system is disordered at
all temperatures.  For $t_{\perp} < t_{\perp}^{\ast}$,
the system becomes ordered below some critical
temperature $T_c$.  At $t_{\perp} = t_{\perp}^{\ast}$
the system has a quantum critical point, at which $T_c
= 0$. Since in sodium vanadate the typical values of
both the Ising couplings and the hopping amplitude
$t_{\perp}$ are of the order of $0.5$eV, whereas $T_c$
is only $34$K, this material has to be very close to
the quantum critical point.

The proximity to the quantum critical point may be
responsible for the strong dependence of $T_c$ on the
hydrostatic pressure observed in sodium vanadate
\cite{Nakao,Loa}. We note that the fact that the
transition temperature is suppressed by pressure can be
naturally explained in our model. Both the hopping
amplitude $t_{\perp}$ and the amplitudes of the dipolar
interactions between the rungs, $V_{\bf n m}$, increase
under pressure. However, $t_{\perp}$, which involves
the overlap of electronic wave functions, depends on
the distances between ions more strongly then $V_{\bf
nm}$. Since the hopping amplitude tends to disorder the
system, the transition temperature decreases under
pressure.

The isospin Hamiltonian Eq.(\ref{HT}) is the
Hamiltonian of the Ising model with long-range
interactions in a transverse field. Recently, a
considerable progress was made in the calculation of
the linear response functions of such systems, which
close to a quantum critical point, become to a large
extent \cite{Sachdev}. While those results could be
useful to describe the response function at wave
vectors ${\bf q}$ close to ${\bf Q}$ ({\em e.g.}, the
temperature dependence of the shape of neutron
diffraction peaks), in this paper we are mostly
interested in the dielectric and magnetic
susceptibility of sodium vanadate at ${\bf q} = 0$, as
the latter is measured in optical experiments. We use
the random phase approximation (RPA) to obtain the
necessary correlation functions of the isospin
operators. Then, for our purposes, it is sufficient to
describe the ordered state in the mean field
approximation.

\begin{figure}
\centering
\includegraphics[width=6cm]{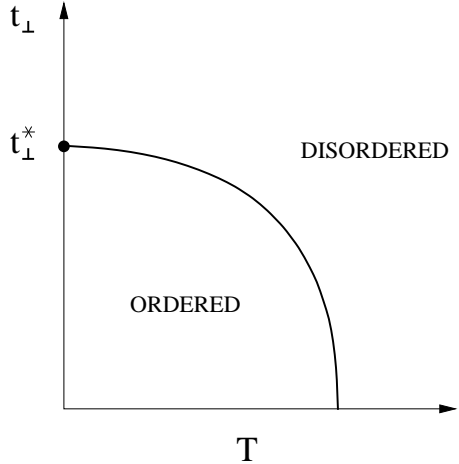}
\caption{
\footnotesize
Generic phase diagram of the model described by Eq.(\ref{HT}) in
the $T$-$t_{\perp}$ plane.}
\label{diagram}
\end{figure}

In this approximation $M_{\bf n}^x$ and $M_{\bf n}^z$
are found from
\begin{equation}
\left\{
\begin{array}{l}
2t_{\perp} M_{\bf n}^z + M_{\bf n}^x \sum_{\bf m}
V_{\bf nm} M_{\bf m}^z = 0,\\ \\ M_{\bf n} =
\sqrt{\left(M_{\bf n}^x\right)^2 + \left(M_{\bf
n}^z\right)^2} = \frac{1}{2} \tanh \frac{H_{\bf
n}}{2T}.
\end{array}
\right.
\label{MFeq}
\end{equation}
The first equation fixes the orientation of the vector
$\langle {\bf T}_{\bf n} \rangle$, which has to be
parallel to the vector of the ``field'' applied to the
isospin at the site ${\bf n}$, ${\bf H}_{\bf n} =
(2t_{\perp}, 0, - \sum_{\bf m} V_{\bf nm} M_{\bf
m}^z)$.  The second equation gives the length of
$\langle {\bf T}_{\bf n} \rangle$ (in that equation
$H_{\bf n} = | {\bf H}_{\bf n}|$).

Above $T_c$ we have
\begin{equation}
\left\{
\begin{array}{lcl}
M_{\bf n}^x & = & \frac{1}{2} \tanh
\frac{t_{\perp}}{T},\\ \\ M_{\bf n}^z & = & 0,
\end{array}
\right.
\label{Mabove}
\end{equation}
while for the ordered states shown in
Fig.~\ref{costates}, the order parameter is given by
Eq.(\ref{odpar}) and $M^x_{\bf n}$ is constant below
$T_c$:
\begin{equation}
M^x_{\bf n} = M^x = \frac{2 t_{\perp}}{|V({\bf Q})|}.
\label{Mxbelow}
\end{equation}
The absolute value of the magnetization, $M =
\sqrt{(M^x)^2 + (M^z)^2}$, satisfies
\begin{equation}
M = \frac{1}{2} \tanh \frac{
\sqrt{\left(2t_{\perp}\right)^2 + \left(M^z V({\bf
Q})\right)^2}} {2 T}, \label{Mbelow}
\end{equation}
where $V({\bf Q})$ is the value of the Fourier
transform of $V_{\bf nm}$ at ${\bf q} = {\bf Q}$:
\begin{equation}
V({\bf Q}) = \sum_{{\bf n}} e^{-i{\bf Q}\cdot({\bf x_n} - {\bf
x_m})}
V_{\bf n m}.
\end{equation}

Finally, at the critical temperature
\begin{equation}
2 t_{\perp} + M^x(T_c) V({\bf Q}) = 0,
\label{Tceq}
\end{equation}
where $M^x(T)$ is given by Eq.(\ref{Mabove}).  Clearly, $V({\bf
Q})$ has to be negative and $V({\bf q})$ has to have a minimum
at ${\bf q} = {\bf Q}$ (the location of the minimum determines
the type of the charge-ordered structure).  Furthermore,
Eq.(\ref{Tceq}) only has a solution for $|V({\bf Q})| \geq 4
t_{\perp}$.  In other words, in the mean field approximation,
the critical point is reached at $t_{\perp}^{\ast} =
\frac{|V({\bf Q})|}{4}$.

\section{Isospin excitations}
\label{excitons}

The isospin excitations described by the Hamiltonian
(\ref{HT}) are the rung excitons, {\em i.e.} the
electronic excitations from the symmetric state with
energy $-t_{\perp}$ to the antisymmetric state with
energy $+t_{\perp}$, which can propagate from rung to
rung due to the Coulomb interactions between the
electrons. This can be most easily understood by
performing a rotation in the isospin space around the
$y$-axis, under which $T^{x} \rightarrow T^{z}$ and
$T^{z} \rightarrow -T^{x}$ (see
Appendix~\ref{AppendixA}). Then the excitation from the
symmetric to the antisymmetric electron state on a rung
corresponds to a spin flip. Due to the Ising
interaction between the isospins on different rungs,
which in the new basis has the form, $\frac{1}{2}
\sum_{\bf nm} V_{\bf nm} \, T^x_{n} \, T^x_{m}$, the
spin flip can hop from rung to rung.

The dispersion of these excitations can be found, {\em
e.g.}, by considering the retarded Green function
\begin{equation}
\langle\langle T^z | T^z \rangle\rangle_{\omega,{\bf
q}} = -i \int_0^{\infty}\!\!dt e^{i (\omega + i \delta)
t} \sum_{\bf n} e^{-i {\bf q} \cdot {\bf x}_{n}}
\left\langle \left[ T^z(t), T^z(0) \right]
\right\rangle
\end{equation}
(here we use the original basis in the isospace). This
Green function is calculated in the random phase
approximation (RPA) in Appendix~\ref{AppendixA} (see
Eqs.~(\ref{greenabove}) and (\ref{greenbelow})). From
the poles of the RPA Green function we obtain the
energy of the isospin excitation with wave vector ${\bf
q}$, valid both above and below $T_c$:
\begin{equation}
E_{\bf q} = \sqrt{(2 t_{\perp})^2 + 2 t_{\perp} M^x  V({\bf q})
+ \left(|V({\bf Q})| M^z\right)^2},
\label{E(q)}
\end{equation}
where the averages $M^x$ and $M^z$ are given by
Eq.(\ref{Mabove}) above $T_c$ and Eqs.(\ref{Mxbelow}) and
(\ref{Mbelow}) below $T_c$.

Since the energy of the electron excitation on a rung,
$2 t_{\perp} \sim 0.7$eV \cite{Smolinski}, is rather
large, the isospin excitations are irrelevant for the
low-temperature properties of sodium vanadate, unless
this large energy is compensated by a large band width
(the latter is determined by the amplitude of the
variation of $V(\bf q)$, which for sodium vanadate is
also expected to be of the order of $1$eV).  We thus
assume, that close to the bottom of the exciton band
({\em i.e.}, for ${\bf q} \sim {\bf Q})$, the exciton
energy is comparable to temperature (or smaller).

The existence of soft isospin excitations implies a
strong temperature dependence of the exciton gap,
$\Delta = E_{\bf Q}$. The typical temperature
dependence of the isospin gap is shown in
Fig.~\ref{Delta}. Above $T_c$ the ``transverse
magnetization'' $M^x$ grows when temperature decreases
(see Eq.(\ref{Mabove})), which results in a decrease of
the gap (we recall that $V({\bf Q})$ in Eq.(\ref{E(q)})
is negative). At the transition temperature, $T_c$, the
exciton gap disappears in accordance with
Eq.(\ref{Tceq}). Below $T_c$, the order parameter,
$M^z$, increases as temperature decreases, while $M^x$
stays constant (see Eqs. (\ref{Mxbelow}) and
(\ref{Mbelow})).  As a result the exciton gap again
reappears.

\begin{figure}
\centering
\includegraphics[width=7cm]{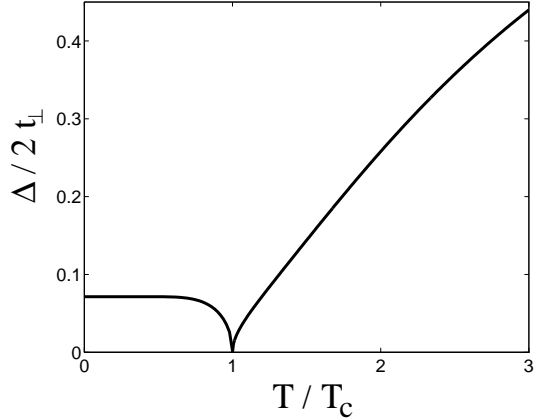}
\caption{ \footnotesize Temperature dependence of the
isospin gap $\Delta$. The value of $|V({\bf Q})|$, used
for this calculation, is close to its quantum critical
value, $2 t_{\perp}$, so that the isospin gap $\Delta
\ll 2t_{\perp}$.} \label{Delta}
\end{figure}

\section{Anomaly of dielectric susceptibility at $T_c$}
\label{anomaly}

The isospin excitations discussed in the previous section can be
excited by applying an electric field in the direction of the
rungs
($a$-direction).  Within the isospin model the interaction with
such a field, $E_a(t)$, is described by the Hamiltonian
\begin{equation}
H_{int} = e l E_a (t) \sum_{\bf n} T^z_{\bf n},
\label{Hint}
\end{equation}
where $l \approx \frac{a}{3}$ is the length of the
rung.  Then the contribution to the dielectric
susceptibility of the electrons occupying the $d_{xy}$
orbitals on the vanadium sites is:
\begin{equation}
\chi_a(\omega,{\bf q}) = - C
\langle\langle T^{z}| T^{z}\rangle\rangle_{\omega,{\bf q}},
\label{chi}
\end{equation}
where the constant $C = \frac{2 e^2 l^2}{abc}$, $a$,
$b$, and $c$ being the lattice constants of the
high-temperature unit cell, and the retarded Green
function $\langle\langle T^z | T^z
\rangle\rangle_{\omega,{\bf q}}$ above and below $T_c$
is given by Eqs.~(\ref{greenabove}) and
(\ref{greenbelow}), respectively. Using these equations
we obtain
\begin{equation}
\chi_a(\omega,{\bf q}) = - C \frac{2 t_{\perp}
M^x}{\left(\omega^2 - {E_{\bf q}}^2 + i \delta
\mbox{sign} \omega \right)} \label{chiabove}
\end{equation}
for the charge susceptibility above $T_c$ (here, $M^x$
is given by Eq.(\ref{Mabove})) and
\begin{equation}
\chi_a(\omega,{\bf q}) = - C
\frac{(2 t_{\perp})^2}{|V({\bf Q})|} \frac{1}
{\left(\omega^2 - {E_{\bf q}}^2 + i \delta \mbox{sign} \omega
\right)}
\label{chibelow}
\end{equation}
below $T_c$.

The frequency dependence of the dielectric
susceptibility was measured by optical absorption
\cite{Damascelli,Popova99,Popova97}. Its imaginary part
shows a rather broad peak at energy $\sim 1$eV, which
was assigned in Ref.~\cite{Damascelli} to
bonding-antibonding excitations of electrons on rungs.
According to Eq.(\ref{chiabove}) the peak should be
located at the energy of the isospin excitations with
${\bf q} = 0$. As for the zigzag structure, shown in
Fig.~4a, this excitation lies far from the bottom of
the exciton band, located at ${\bf Q} =
(\frac{\pi}{f_1},\frac{\pi}{f_2})$, its energy,
$E_{{\bf q}=0}$, should indeed be rather large ($\sim 2
t_\perp$).  On the other hand, for the chain structure
(see Fig.~4b) the bottom of the exciton band would be
at ${\bf q} = 0$ and the soft isospin excitations would
result in a strong peak in the optical absorption at
low energy, $E_{{\bf q}=0}\approx 0$, which is not
observed in these experiments.

The type of charge ordering also determines the type of
the anomaly in the temperature dependence of the static
dielectric susceptibility close to $T_c$. Such an
anomaly appears due to the redistribution of charge at
the phase transition. The order parameter $M^z$, as
well as $M^x$ and the isospin energy $E_{\bf q}$ depend
on $T$, which makes the dielectric susceptibility
temperature-dependent (see Eqs.(\ref{chiabove}) and
(\ref{chibelow})).

For the antiferroelectric ordering of charges, as in
the case of the zigzag ordering (see
Fig.~\ref{costates}a), the susceptibility to the
uniform field is finite at $T_c$ (see
Eq.(\ref{chiabove}) with $E_{{\bf q}=0}$ finite), while
its temperature derivative has a discontinuity at the
transition temperature, as is shown in
Fig.~\ref{chistat}.  (This is analogous to the anomaly
in the magnetic susceptibility of the Ising
antiferromagnet). On the other hand, for the
ferroelectric type of ordering, such as the chain-like
structure shown on Fig.~\ref{costates}b, the
susceptibility would diverge at $T_c$.  The anomaly in
the temperature dependence of the dielectric
susceptibility of NaV$_2$O$_5$ has been measured at
microwave and far-infrared frequencies
\cite{Smirnov,Sekine,Poirier}}. The observed anomaly is
similar to that shown in Fig.~\ref{chistat} and
consistent with a charge ordering of the
antiferroelectric type \cite{Smirnov}, although above
$T_c$ the dielectric susceptibility of sodium vanadate
continues to grow with temperature, whereas the RPA
predicts a decrease.

\begin{figure}
\centering
\includegraphics[width=7cm]{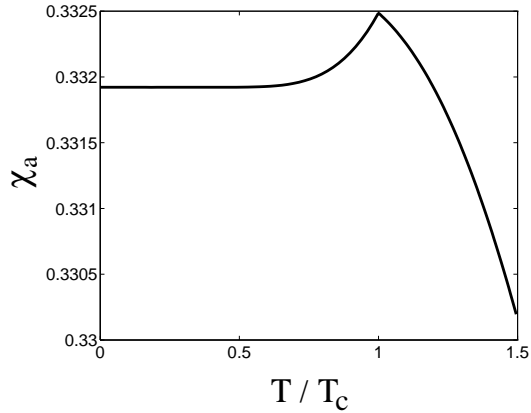}
\caption{ \footnotesize The typical temperature
dependence of the static dielectric susceptibility to
electric fields applied in the $a$-direction calculated
in the RPA approximation. The parameters of the isospin
Hamiltonian were chosen such that the value of $T_c$ is
much smaller than $t_{\perp}$.} \label{chistat}
\end{figure}

\section{Optical absorption at low energies}
\label{oabsorption}

\begin{figure}
\centering
\includegraphics[width=7cm]{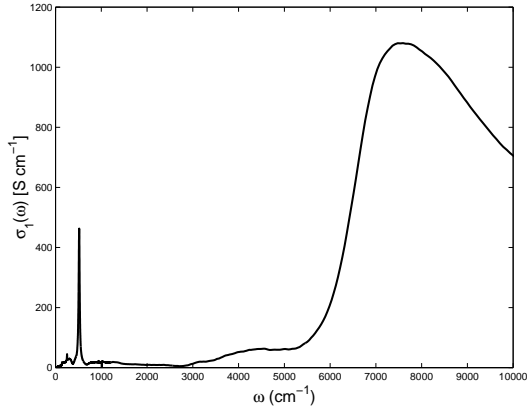}
\caption{ \footnotesize The frequency dependence of the
real part of the optical conductivity of NaV$_2$O$_5$
at $300$K in the midinfrared range (courtesy of Prof.
D. van der Marel \cite{Damascelli}).}\label{Dirk}
\end{figure}

A remarkable feature of the optical absorption spectrum
of sodium vanadate at low-energies is that the
low-energy shoulder of the peak at $8000$cm$^{-1}$
extends all the way down to $\sim3000$cm$^{-1}$ (see
Fig.~\ref{Dirk}). Furthermore, at lower frequencies
there is another broad absorption band stretching from
$\sim100$cm$^{-1}$ up to $\sim1500-2000$cm$^{-1}$ with
a maximum at $\sim 300$cm$^{-1}$
\cite{Damascelli,Popova99}. A very broad peak with a
maximum at $600$cm$^{-1}$ is also observed in Raman
scattering \cite{Golubchik,Popova99,Fisher}. The large
width of these peaks, as well as a number of the Fano
resonances observed in the far-infrared absorption
spectrum \cite{Popova99,Damascelli}, indicate the
presence of a broad continuum of low-energy excitations
in sodium vanadate, which covers, practically, the
whole midinfrared range of frequencies.

The low-energy continuum, observed both above and below
$T_c$, can, in principle, be due to the photoexcitation
of several isospin excitations lying close to the
bottom of the exciton band. To describe such processes
one has to go beyond the RPA. Then the one-exciton
states are mixed with the states containing three,
five, and more excitons. However, since the number of
excitons created in the photoabsorption is odd, the
momentum conservation does not allow all of them
simultaneously to have low energy. Thus, the pure
exciton continuum should begin at rather high energies
and cannot account for the mid-infrared continuum in
sodium vanadate.

To substantiate the point that the isospin excitations
alone cannot give rise to the observed low-energy
optical absorption, we consider an isolated vanadium
ladder with the isospin Hamiltonian:
\[
H_T = - 2t_{\perp}\sum_n T_n^x + V \sum_{n} T_n^z
T_{n+1}^z.
\]
For open boundary conditions, this Hamiltonian can be
diagonalized exactly. To this end we first perform a
rotation around the $y$-axis in the isospin space ({\em
cf.} Appendix). The resulting Hamiltonian is
diagonalized by transforming the isospins to fermions
(using the Jordan-Wigner transformation) and then
applying the Bogoliubov transformation \cite{LSM}. The
exact dispersion of the fermionic excitations reads:
\[
E_q = \sqrt{\left(2t_{\perp}+\frac{V}{2}\cos q\right)^2
+ \left(\frac{V}{2}\sin q\right)^2},
\]
with $q$ a quantum number that follows from a
transcendental equation \cite{LSM}, and which lies in
the interval $0 < q < \pi$. Thus, the isospin gap is
given by
\[
\Delta = E_{\pi} = \left| 2 t_{\perp} - \frac{V}{2}
\right|.
\]

\begin{figure}
\centering
\includegraphics[width=7cm]{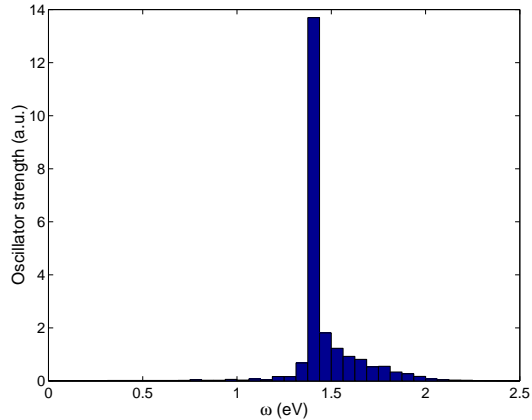}
\caption{ \footnotesize The oscillator strength for an
isolated two-leg ladder of $N = 70$ rungs with very
small value of the isospin gap at $q = \pi$ (see
explanations in the text). The narrow peak is due to
the single exciton absorption. The absorption above the
main peak is due to the photoexcitation of three
excitons. The very small oscillator strength for the
frequencies below the frequency of the main peak is due
to the one-exciton absorption and is a finite-size
effect.} \label{3ex}
\end{figure}

The oscillator strength of a $n$-exciton state ($n$
odd) may now be expressed in terms of the ground state
expectation value of strings of Fermi operators. The
latter may be evaluated using Wick's theorem which
results in determinants of matrices with components
that follow from the Bogoliubov transformation
coefficients. For the $1$- and $3$-exciton states, this
calculation has been performed explicitly in
Ref.~\cite{BK}. Using these results, we find for a
ladder of $N=70$ rungs and  $2 t_{\perp} = 0.7$eV the
absorption spectrum plotted in Fig.~\ref{3ex}. As
NaV$_2$O$_5$ seems to be close to the quantum critical
point, $2 t_{\perp} = \frac{V}{2}$, (see discussion in
Sec.~3), we choose $\frac{V}{2} = 0.6999$eV, which
gives an isospin gap $10^{-4}$eV. The narrow peak is
due to the single-exciton absorption. The absorption
above the main peak is due to the photoexcitation of
three excitons. This process contributes to the broad
high-energy wing of the main absorption peak observed
in sodium vanadate \cite{Damascelli}. The small
oscillator strength at frequencies below the frequency
of the main peak is due to the one-exciton absorption
and is a finite-size effect (breaking of the momentum
selection rule). For small photon frequencies the
absorption due to the $5$-exciton states is negligible
due to the small phase volume of these states.

\begin{figure}
\centering
\includegraphics[width=6cm]{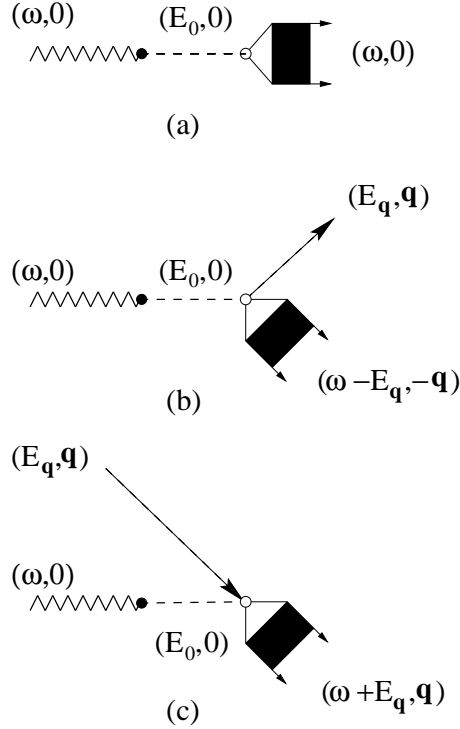}
\caption{ \footnotesize Diagrams describing: (a) the
``charged magnon'' mechanism, (b) the photoexcitation
of the two-spinon continuum and a low-energy exciton,
and (c) the photoexcitation of the two-spinon continuum
with the annihilation of a low-energy exciton. The wavy
and dashed lines correspond to, respectively, a photon
and a charge exciton, while the black square indicates
the two-spinon continuum.} \label{diagrams}
\end{figure}

In Ref.~\cite{Damascelli} the low-energy optical was
associated with the two-spinon continuum. Since
magnetic excitations cannot be directly induced by
photon absorption, the authors of
Ref.~\cite{Damascelli} invoked the so-called ``charged
magnon'' mechanism. The corresponding Feynman diagram
is shown in Fig.\ref{diagrams}(a): The process involves
the creation of a virtual high-energy isospin
excitation with ${\bf q} = 0$, which then decays into
two spinons. However, the coupling of the isospin
excitation with zero wave vector to spinons is only
nonzero if the system has a permanent electric dipole.
In other words, the magnons have charge only if sodium
vanadate is ferroelectric, which would require the
chain-like electronic ordering (cf. Fig~4b) both above
and below $T_c$. For a long time sodium vanadate was,
indeed, believed to have such a chain structure
\cite{Carpy}. However, this structure was found to be
inconsistent (certainly, in the high-temperature phase)
with recent experimental data (see discussion in Sec.
5).

Thus the coupling of a photon with zero wave vector to
two spinons is zero, and we have to find another
mechanism for the low-energy optical absorption. In the
remainder of this section, we shall show that the
low-energy absorption may result from the
photoexcitation of a three-particle continuum: two
spinons and one low-energy isospin excitation.  This
mechanism is illustrated by the diagram in
Fig.\ref{diagrams}(b). This process also takes place
via a virtual isospin excitation with zero wave vector
and high energy $E_0$, which then decays into a
low-energy isospin excitation with the wave vector
${\bf q}$ close to the bottom of the band, ${\bf q}
\sim {\bf Q}$, and two spin excitations, which carry
the momentum $-{\bf q}$. Alternatively, the thermally
excited isospin excitation can be annihilated in the
process of optical absorption (see
Fig.\ref{diagrams}(c)).

Our mechanism requires the coupling of {\em two}
isospin excitations (one of high energy, another of low
energy) to two spin excitations. In sodium vanadate
such a coupling results from the fact that the spin
exchange of electrons on neighboring rungs of a two-leg
ladder cannot be separated from the exchange of their
isospins. The Hamiltonian describing the spin-isospin
coupling was discussed in Sec.~\ref{spin-isospin}. As
an exact treatment of the spin-isospin model is not
possible we use a number of approximations to calculate
the optical absorption spectrum. First we note that the
Hamiltonian of the spin-isospin interaction given by
Eqs.~(\ref{HST1}), (\ref{HST2}), and (\ref{HST3}) can
be written in the form:
\[
H_{ST} = H_{ST}^{(1)} + H_{ST}^{(2)} + H_{ST}^{(3)} =
\sum_{\bf n} {\cal T}_{\bf n} {\cal S}_{\bf n},
\]
where ${\cal T}_{\bf n}$ contains all isospin
operators:
\begin{equation}\label{calT}
{\cal T}_{\bf n} = A\left(T^x_{\bf n} T^x_{{\bf n}+{\bf
f}_1} + T^y_{\bf n} T^y_{{\bf n}+{\bf f}_1}\right) +
A^{\prime} T^z_{\bf n} T^z_{{\bf n}+{\bf f}_1} + B
\left(T^z_{{\bf n}+{\bf f}_2} - T^z_{{\bf n}+{\bf
f}_1-{\bf f}_2}\right),
\end{equation}
and
\[
{\cal S}_{\bf n} = {\bf S}_{\bf n} \!\cdot\! {\bf
S}_{{\bf n} + {\bf f}_1}.
\]

\begin{figure}
\centering
\includegraphics[width=5cm]{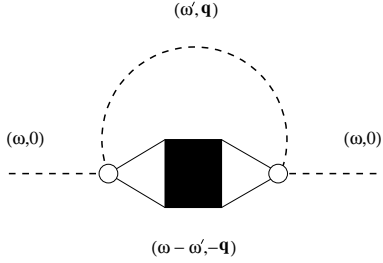}
\caption{ \footnotesize The diagram describing the
correction to the Green function of the isospin
excitation due to scattering on the spin excitations.
This correction which gives rise to low-energy optical
absorption.} \label{correction}
\end{figure}

Although in our model the spin dynamics is inseparable
from the dynamics of the isospins,  we may, however,
obtain a ``pure spin Hamiltonian'' by treating the
isospin operators in $H_{ST}$ within the mean-field
approximation:
\begin{equation} H_{S} = \sum_{\bf n}\!
J_{{\bf n}, {\bf n} + {\bf f}_1} \left({\bf S}_{\bf n} \!\cdot\!
{\bf S}_{{\bf n} + {\bf f}_1}\right),
\end{equation}
where the effective spin-exchange constant is obtained
by the thermal average of ${\cal T}_{\bf n}$ over the
state of the isospin system:
\[
J_{{\bf n}, {\bf n} + {\bf f}_1} = \langle {\cal T}_{\bf n}
\rangle_{H_T}.
\]
As was already discussed in Sec.~\ref{spin-isospin},
the thus obtained spin-exchange constant depends on
temperature and becomes alternating along the spin
chains below $T_c$. Since according to our assumption
the charge ordering is mainly driven by the pure
isospin Hamiltonian $H_T$, we neglect the effect of the
spin-isospin interactions on the state of the isospin
system. The remaining part of the spin-isospin
Hamiltonian (the residual interaction) describes the
scattering of isospin excitations on spin excitations.
This scattering gives rise to the low-energy absorption
and will be treated as a perturbation.

In this paper we consider the optical absorption only
above $T_c$ ({\em i.e.}, no charge ordering and no spin
gap). Furthermore, the isospin excitations are treated
as bosons. In particular, the operator
\[
T^z_{\bf q} = \frac{1}{\sqrt{N}}\sum_{\bf n} e^{-i{\bf q} {\bf
x}_{\bf n}} T^z_{\bf n}
\]
annihilates the isospin excitation with the wave vector ${\bf q}$
and creates such an excitation with the wave vector $-{\bf q}$:
\[
T^z_{\bf q} \sim
\sqrt{\frac{t_{\perp}M}{E_{\bf q}}}
\left(b_{- {\bf q}}^{\dagger} + b_{\bf q}\right).
\]
The coefficient in the right-hand side of the last
equation was found by comparison of this equation with the Green
function Eq.(\ref{greenabove}).

Since the operator ${\cal T}_{\bf n}$ (see
Eq.(\ref{calT})) contains products of two isospin
operators, it either results in the creation and
annihilation of two isospin excitations, or in the
scattering of the isospin excitation. In particular, if
one of the isospin excitations has zero wave vector (as
is the case in the optical absorption), then
\[
{\cal T}_{\bf q} \sim g_{\bf q} \frac{t_{\perp}M}{\sqrt{E_0 E_{\bf
q}}} \left(b_{- {\bf q}}^{\dagger} + b_{\bf q}\right)
\left(b_0^{\dagger} + b_0\right),
\]
where $g_{\bf q}$ is the coupling constant.

To lowest order in the residual spin-isospin
interaction, the Green function of the isospin
excitation obtains the correction shown in
Fig.~\ref{correction}, which leads to the low-energy
optical absorption described by the diagrams in
Figs.~\ref{diagrams} (b) and (c). The corresponding
contribution to the imaginary part of the dielectric
susceptibility is:
\begin{eqnarray}\label{answer}
\Delta \chi_a^{\prime\prime}(\omega,0) &=& \frac{K}{N}
\sum_{\bf q} \frac{|g_{\bf q}|^2}{E_0 E_{\bf q}} \Bigg[
\chi_S^{\prime\prime}(\omega-E_{\bf q},-{\bf q}) \left(
\coth \frac{E_{\bf q}}{2T} - \coth\frac{(E_{\bf q} -
\omega)}{2T}\right) \nonumber\\ &\!\!\!+\!\!\!&
\chi_S^{\prime\prime}(\omega+E_{\bf q},{\bf q}) \left(
\coth\frac{E_{\bf q}}{2T} - \coth \frac{(E_{\bf q}
+\omega)}{2T}\right) \Bigg].
\end{eqnarray}
Here, the first term in square brackets is due to
photoexcitation of one low-energy exciton plus two
spinons (see Fig.~\ref{diagrams}b), while the second
term describes annihilation of a thermally excited
low-energy excitation, accompanied by the creation of
two spin excitations (see Fig.~\ref{diagrams}c). In
Eq.(\ref{answer}) we have defined
\[
K = \frac{1}{abc} \left(\frac{2elt_{\perp}^2M^2}{E_0^2
- \omega^2}\right)^2,
\]
while $\chi_S(\omega,{\bf q}) = \chi_S(\omega,q_1)$ is
the susceptibility of the Heisenberg spin-$\frac{1}{2}$
chain:
\begin{equation}\label{chiSdef}
 \chi_S(\omega,q_1) = i\int_0^{\infty}dte^{i\omega t-iq_1n}
 \langle \left[{\cal S}_n(t),{\cal S}_0(0)\right] \rangle_{H_S}
\end{equation}
(here $q_1$ is the wave vector in the chain direction
measured in units of $\frac{1}{b}$). Since no exact
expression for this susceptibility is known that holds
for all values of $q$, $\omega$, and $T$, we calculate
it by substituting the Heisenberg model by the
renormalized XY model, which is equivalent to a
half-filled chain of spinless fermions with the
dispersion $\varepsilon(k) = - p J \cos k $, where $p =
1 + 2/\pi$ \cite{Bulaevskii}. Then
\[
{\cal S}_n \rightarrow -\frac{p}{2}\left(c_{n+1}^{\dagger} c_n +
c_n^{\dagger} c_{n+1} \right),
\]
and
\[
\chi^{\prime\prime}_S(\omega,q) = \frac{p^2}{4} \int_{-\pi}^{\pi}
d k \delta\left(\omega + \varepsilon_k - \varepsilon_{k+q}\right)
\cos^2 \left(k + \frac{q}{2}\right)
\left[\tanh\frac{\varepsilon_{k+q}}{2T} -
\tanh\frac{\varepsilon_{k}}{2T}\right].
\]
According to our assumption, the isospin excitations
only are soft in a small vicinity of $q_1 = \pi$. Then
the wave vector of the spin excitations in the
low-energy continuum is also close to $\pi$, in which
case the analytical expression for the imaginary part
of the spin susceptibility is:
\begin{equation}\label{chiSXY}
  \chi_S^{\prime\prime}(\omega,\pi) = \theta(\Omega^2 - \omega^2)
  \frac{p^2}{\Omega}
  \sqrt{1 - \left(\frac{\omega}{\Omega}\right)^2}
  \tanh \frac{\omega}{4T}.
\end{equation}
Here, $\Omega = 2 p J$ is the bandwidth of the spinless
fermions, which corresponds to the maximal energy of
the two-spinon excitation in the Heisenberg model.

\begin{figure}
\centering
\includegraphics[width=7cm]{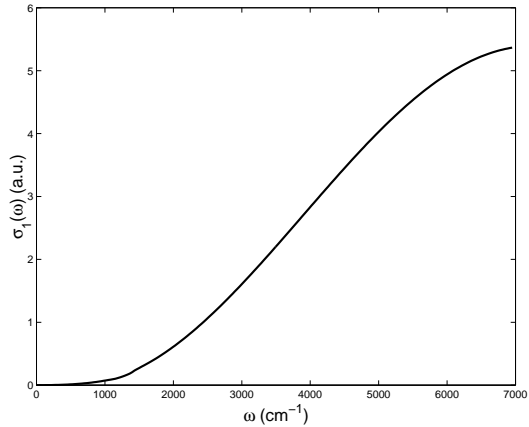}
\caption{ \footnotesize The low-energy part of the
optical absorption spectrum due to the process shown in
Figs.\ref{diagrams}(b) and (c).} \label{absorption}
\end{figure}

In Fig.~\ref{absorption}, we show the real part of
optical conductivity, $\sigma_1(\omega) \propto \omega
\chi_a^{\prime\prime}(\omega,0)$, resulting from Eqs.
(\ref{answer}) and (\ref{chiSXY}) for the model
parameters: $\Delta = 200$K, $T = 300$K, and $J = 560$K
is shown Fig.~\ref{absorption}. The spectrum has two
singularities at $\omega = \Omega \pm \Delta$ due to
the fact that the maximal energy carried by two spin
excitations cannot exceed $\Omega$. The contribution of
the process with the annihilation of an isospin
excitation (see Fig.~\ref{diagrams}c) has its maximum
at $\omega \sim \Omega$ and decreases exponentially (as
$e^{-\beta \omega}$) at larger frequencies. Therefore,
the absorption at $\Omega,T \ll \omega \ll E_0$ is
entirely due to the process shown in
Fig.~\ref{diagrams}b. At these frequencies the optical
absorption spectrum measures, basically, the density of
isospin excitations. Since the spectrum of these
excitations stretches from very low energies near the
bottom of the band at ${\bf q} = {\bf Q}$ up to the
energies of the order of $2 t_{\perp}$, a (relatively
small) optical absorption due to the photoexcitation of
the three-particle continuum should occur at all
frequencies between $\Delta$ and $E_0$. This explains
the midinfrared absorption continuum observed in
NaV$_2$O$_5$ and described in the beginning of this
section. We note that, in agreement with our theory,
this continuum is only observed when the electric field
is directed along the ladder rungs (along ${\bf a}$).

To end this section, we point out, that our mechanism
of the optical absorption at low photon energies is
very similar to the Lorenzana-Sawatzky mechanism of the
phonon-mediated photoexcitation of two magnons
\cite{LS}. The difference lies in the fact that the
phonon dispersion is usually small or comparable to
that of magnons, so that the shape of the
three-particle continuum is mainly determined by the
spin excitations \cite{Nagaosa}. On the other hand, in
our case the shape of the optical spectrum is governed
by the isospin excitations, which a have much wider
band than spinons.

\section{Discussion and Conclusions}
\label{disclusions}

In this paper we discussed the spin-isospin model of
sodium vanadate. We showed that when the rungs of the
vanadium two-leg ladders are occupied mostly by a
single electron, the spectrum of low-energy excitations
may include, not only spin, but also charge
excitations. The latter are the bonding-antibonding
electronic excitations on the ladder rungs, which
become itinerant due to the Coulomb interactions
between electrons on different rungs.

We described the dynamics of these excitations with
isospin-$\frac{1}{2}$ operators. The isospin
Hamiltonian is identical to the Hamiltonian of the
Ising model in a transverse field. In this model one
obtains a transition into an ordered state al low
temperatures, which corresponds to the charge ordering
of electrons on rungs of the vanadium ladders. At the
transition temperature the spectrum of the isospin
excitations becomes gapless.

Furthermore, we showed that the dynamics of the spins
and isospins are coupled to each other. A pure spin
exchange does not exist - the exchange of spins
necessarily involves the exchange of isospins. As a
result the effective spin exchange constants depend on
temperature. Below the charge ordering temperature the
spin exchange constants alternate along ladders, which
opens a spin gap.

As we have shown, the spin-isospin Hamiltonian explains
the anomaly in the dielectric susceptibility at $T_c$
and the optical absorption at low frequencies, observed
in sodium vanadate. The shape of the anomaly in the
dielectric constant proves that the charge ordering
below $T_c$ in NaV$_2$O$_5$ is of an antiferroelectric
(zigzag) type.

A considerable part of this paper was devoted to the
calculation of the optical absorption within the
spin-isospin model. We argued that neither pure isospin
excitations, nor pure spin excitations can explain the
broad bands in the optical absorption spectrum of
sodium vanadate found in the whole range of
mid-infrared frequencies. We showed that this
absorption can be due to the photoexcitation of a
three-particle continuum (one low-energy exciton and
two spinons).  We also note that below $T_c$, the
quasi-momentum conservation allows for the optical
excitation of a single low-energy excitation, as the
increase of the lattice period folds these modes into
the zone center. Indeed, both in optical absorption
\cite{PopovaLT} and Raman spectra \cite{Fisher},
several lines have been observed whose intensity shows
an anomalous temperature dependence.

Thus the spin-isospin model allows us to explain
qualitatively the main experimental facts on sodium
vanadate, {\em i.e.}, the charge ordering, the spin-gap
opening, and the low-energy absorption. The main
difficulty of this model is to explain the low value of
the charge ordering temperature. One possibility
mentioned above is the proximity to the quantum
critical point, which, however, requires a fine tuning
of the model parameters (the transverse field and the
Ising interaction between the isospins). Another
possibility, the frustration of the Ising interactions
due to the relative shifts between neighboring ladders,
will be discussed in detail elsewhere.

In conclusion, in sodium vanadate the electronic charge
dynamics seems to play an important role, and the
coupling between the charge and spin degrees of freedom
is crucial in understanding the properties of this
material. Finally, we note that a similar spin-isospin
model could be used to describe the interplay between
the charge and antiferromagnetic ordering in the
quasi-two-dimensional organic system
$\kappa$-(BEDT-TTF)$_2$X (see, {\em e.g.},
\cite{Fukukino,Ross}), in which the role of the V rungs
is played by dimers of BEDT-TTF molecules.

This work is supported by the ``Stichting voor
Fundamenteel Onderzoek der Materie (FOM)'' and the
MCS$+$ programme. The authors are grateful to B.
B\"uchner, C. Gros, W. Kremer, P. van Loosdrecht, D.
van der Marel, T. Palstra, and N. Prokof'ev for
fruitful discussions.

\appendix

\section{Spectrum of isospin excitations in RPA}
\label{AppendixA}

In this Appendix we find the spectrum of isospin excitations
both in the disordered and ordered phases.  Above $T_c$, $M^z_{\bf n}
= 0$ and $M^x_{\bf n} = M \neq 0$.  It is convenient to perform
a rotation in the isospace:
\begin{equation}
\left[
\begin{array}{c}
T_{\bf n}^x\\ T_{\bf n}^y\\ T_{\bf n}^z
\end{array}
\right]
\rightarrow
\left[
\begin{array}{c}
T_{\bf n}^{\prime x}\\ T_{\bf n}^{\prime y}\\ T_{\bf n}^{\prime
z}
\end{array}
\right]
=
\left[
\begin{array}{c}
-T_{\bf n}^{z}\\ +T_{\bf n}^{y}\\ +T_{\bf n}^{x}
\end{array}
\right]. \label{rotation}
\end{equation}
In the rotated basis $\langle T^{\prime z}_{\bf n} \rangle = M$
and the
Hamiltonian has the form:
\begin{equation}
H_T = - 2 t_{\perp} \sum_{\bf n} T_{\bf n}^{\prime z} +
\frac{1}{2} \sum_{\bf nm} V_{\bf nm} T^{\prime x}_{\bf n}
T^{\prime x}_{\bf m},
\label{rotham}
\end{equation}

We then use the equations of motion for the operators $T_{\bf
n}^{\prime \pm} = T_{\bf n}^{\prime x} \pm i T_{\bf n}^{\prime
y}$:
\begin{equation}
\pm i {\dot T}_{\bf n}^{\prime \pm} = 2 t_{\perp} T_{\bf
n}^{\prime \pm} + \frac{1}{2} \sum_{\bf m} V_{\bf nm} T_{\bf
n}^{\prime z} \left(T_{\bf m}^{\prime +} + T_{\bf m}^{\prime
-}\right),
\end{equation}
in which we substitute $T_{\bf n}^z$ by its average
value, $M$. This approximation, equivalent to the
decoupling of the Green functions containing more than
two isospin operators \cite{Zubarev,Haley}, gives a
closed system of equations,
\begin{equation}
\left\{
\begin{array}{lcc}
\left[\omega - 2 t_{\perp} - \frac{M}{2} V({\bf q})\right]
\langle\langle T^{\prime +}| T^{\prime -}\rangle\rangle
- \frac{M}{2} V({\bf q})
\langle\langle T^{\prime -}| T^{\prime -}\rangle\rangle
& = & 2 M\\ \\
\frac{M}{2} V({\bf q})
\langle\langle T^{\prime +}| T^{\prime -}\rangle\rangle
+ \left[\omega + 2 t_{\perp} + \frac{M}{2} V({\bf q})\right]
\langle\langle T^{\prime -}| T^{\prime -}\rangle\rangle
& = & 0
\end{array}
\right.,
\end{equation}
for the retarded Green functions $\langle\langle A | B
\rangle\rangle \equiv \langle\langle A | B
\rangle\rangle_{\omega, {\bf q}}$, defined by
\begin{equation}
\langle\langle A | B \rangle\rangle_{\omega,{\bf q}} = -i
\int_0^{\infty}\!\!dt
e^{i (\omega + i \delta) t} \sum_{\bf n} e^{-i {\bf q} \cdot {\bf
x}_{n}} \left\langle \left[ A, B \right] \right\rangle.
\end{equation}
In this way we get:
\begin{equation}
\left\{
\begin{array}{lcr}
\langle\langle T^{\prime +}| T^{\prime
-}\rangle\rangle_{\omega,{\bf q}} & = & 2 M
\frac{\left(\omega + 2 t_{\perp} + \frac{M}{2} V({\bf
q})\right)}{\omega^2 - E_{\bf q}^2 +
i \delta \mbox{\scriptsize sign} \omega},\\ \\
\langle\langle T^{\prime -}| T^{\prime
-}\rangle\rangle_{\omega,{\bf q}} & = & -
\frac{M^2 V({\bf q})}
{\omega^2 - E_{\bf q}^2 + i \delta \mbox{\scriptsize sign}
\omega}.
\end{array}
\right.
\label{green1}
\end{equation}
Furthermore,
\begin{equation}
\left\{
\begin{array}{lcr}
\langle\langle T^{\prime -}| T^{\prime
+}\rangle\rangle_{+\omega,{\bf q}} & = &
\langle\langle T^{\prime +}| T^{\prime
-}\rangle\rangle_{-\omega,{\bf q}},\\ \\
\langle\langle T^{\prime +}| T^{\prime
+}\rangle\rangle_{\omega,{\bf q}} & = &
\langle\langle T^{\prime -}| T^{\prime-}
\rangle\rangle_{\omega,{\bf q}}.
\end{array}
\right.
\label{green2}
\end{equation}

The Green functions have poles at (plus/minus) the energy of the
isospin excitation with wave vector ${\bf q}$:
\begin{equation}
E_{\bf q} =
\sqrt{2 t_{\perp} \left(2 t_{\perp} + M V({\bf q}) \right)}.
\label{Eabove}
\end{equation}
The band of isospin excitations has its bottom at the wave vector
${\bf q} = {\bf Q}$, where $V({\bf q})$ has its minimum.

Combining Eqs.(\ref{green1}) and (\ref{green2}) we obtain:
\begin{equation}
\langle\langle T^z | T^z \rangle\rangle_{\omega,{\bf q}} =
\langle\langle T^{\prime x}| T^{\prime x} \rangle\rangle = \frac{2
t_{\perp} M}{\left(\omega^2 - {E_{\bf q}}^2 + i \delta \mbox{sign}
\omega \right)}. \label{greenabove}
\end{equation}

In the ordered phase below $T_c$ we introduce on each site ${\bf
n}$ an angle $\phi_{\bf n}$ by
\begin{equation}
\left\{
\begin{array}{ccl}
M_{\bf n}^x & = & M_{\bf n} \cos \phi_{\bf n}\\
M_{\bf n}^z & = & M_{\bf n} \sin \phi_{\bf n}
\end{array}
\right.,
\end{equation}
where we used the original basis.  Then the first of the
self-consistency equations Eq.(\ref{MFeq}) reads:
\begin{equation}
2t_{\perp} \sin  \phi_{\bf n}
+ \cos \phi_{\bf n} \sum_{\bf m} V_{\bf nm}
M_{\bf m} \sin\phi_{\bf m} = 0.
\label{condition}
\end{equation}

We now chose the basis in the  isospin space, such that the
$z$-axis is oriented along $\langle {\bf T}_{\bf n} \rangle$:
\begin{equation}
\left\{
\begin{array}{l}
{T^{\prime}}_{\bf n}^x = \sin \phi_{\bf n} T_{\bf n}^x - \cos
\phi_{\bf n} T_{\bf n}^z,\\
{T^{\prime}}_{\bf n}^y  =  T_{\bf n}^y,\\
{T^{\prime}}_{\bf n}^z = \cos \phi_{\bf n}
T_{\bf n}^x + \sin \phi_{\bf n} T_{\bf n}^z.\\
\end{array}
\right.
\end{equation}
Then $\langle {T^\prime}_{\bf n}^z \rangle = M_{\bf n}$ and
$\langle {T^\prime}_{\bf n}^x \rangle = \langle {T^\prime}_{\bf
n}^y \rangle = 0$.

Using Eq.(\ref{condition}) the linearized equations of motion for
$T^{\prime \pm}$ can be written in the form:
\begin{equation}
\pm i {\dot T}^{\prime \pm}_{\bf n}
= \frac{2 t_{\perp}}{\cos \phi_{\bf n}}
{T^{\prime}}_{\bf n}^{\pm}
+ \frac{1}{2} M_{\bf n} \cos \phi_{\bf n}
\sum_{\bf m} V_{\bf nm} \cos \phi_{\bf m}
\left({T^{\prime}}_{\bf m}^{+} +
{T^{\prime}}_{\bf m}^{-}\right).
\label{eqmobelow}
\end{equation}

Both for the zigzag and the chain ordering,
\begin{equation}
\phi_{\bf n} = e^{i {\bf Q} \cdot {\bf x}_{\bf n}} \phi,
\end{equation}
and $|\phi_{\bf n}| = \phi$.
Then, Eq.(\ref{condition}) gives $M_{\bf n} = M$ and
\begin{equation}
M =  \frac{2 t_{\perp}}{\cos \phi |V({\bf Q})|},
\end{equation}
({\em cf.} Eq.(\ref{Mxbelow})). Furthermore,
from Eq.(\ref{eqmobelow}) we see that
the Green functions in the ordered phase can be obtained from
those above $T_c$ by substituting $2 t_{\perp}$ by $\frac{2
t_{\perp}}{\cos \phi}$ and $V({\bf q})$ by
$V({\bf q})\cos^2 \phi$,
{\em e.g.},
\begin{equation}
\langle\langle T^{\prime +}| T^{\prime
-}\rangle\rangle_{\omega,{\bf q}} = 2 M
\frac{\left(\omega + \frac{2 t_{\perp}}{\cos
\phi} + \frac{M}{2}
\cos^2\phi V({\bf q})\right)}{\omega^2 - E_{\bf q}^2
+ i \delta \mbox{sign} \omega},
\end{equation}
where the energy of the isospin excitations is given by
\begin{equation}
E_{\bf q} = \frac{2 t_{\perp}}{\cos \phi}
\sqrt{1 - \cos \phi^2 \frac{V({\bf q})}{V({\bf Q})}}.
\label{Ebelow}
\end{equation}
Since below $T_c$, $0 < \cos \phi < 1$, the isospin gap in the
ordered phase, $\Delta = E_{\bf Q} = 2 t_{\perp} \tan \phi > 0$.
The dispersion relations Eq.(\ref{Eabove}) and Eq.(\ref{Ebelow})
can be combined to give Eq.(\ref{E(q)}) valid both above and
below $T_c$.

Using Eq.(\ref{condition}), one can check that the Green
functions of $T^{\prime z}$ and any other operator vanish.  Then,
$\langle\langle T^z | T^z \rangle\rangle_{\omega,{\bf q}} =
\cos^2 \phi \langle\langle T^{\prime x}| T^{\prime x}
\rangle\rangle$, so that below $T_c$,
\begin{equation}
\langle\langle T^z | T^z \rangle\rangle_{\omega,{\bf q}} =
\frac{(2 t_{\perp})^2}{|V({\bf Q})|} \frac{1}
{\left(\omega^2 - {E_{\bf q}}^2 + i \delta \mbox{sign}
\omega \right)}
\label{greenbelow}
\end{equation}

\end{document}